# Value of spatially multiplexed SNSPDs with photon-number-resolving capabilities

Fabio Auriemma[1,*] and Mikkel Ejrnaes[1]

[1] CNR - Institute for SuPerconductors, INnovative materials, and devices (SPIN), Via Campi Flegrei, 34, Pozzuoli (ZIP 80078), Napoli, Italy

We develop a combinatorial and Bayesian framework to quantify the information gained by increasing the local photon-number resolution of multiplexed detection systems based on superconducting nanowire single-photon detectors (SNSPDs). Motivated by recent studies that attribute photon-number resolving capabilities to a single SNSPD overcoming the typical on/off detection mechanism, we generalize the existing formulation for the binary devices by deriving the probability response tensor of multiplexed systems based on SNSPDs with an arbitrary number of resolvable states. We include finite efficiency through binomial loss convolution and, separately, assess the impact of dark counts. From this response tensor we derive a generalized $N$-photon detector fidelity, a Bayesian photon-number reconstruction, and the corresponding reconstruction fidelity. We particularize this formulation to study 2-, 3- and 4-state multiplexed systems with different numbers of detectors and show that increasing local photon-number resolution systematically improves both detector fidelity and Bayesian photon-number reconstruction. This substantially relaxes the quadratic scaling between detector number and resolvable photon-number required by purely binary multiplexed arrays. These results provide a quantitative value-of-information assessment for multi-state photon-number-resolving detector arrays.

## I. INTRODUCTION.

Photon-number resolution (PNR) is often a central requirement in quantum optics and quantum information research, with applications that include linear-optical quantum computing, quantum communication, source characterization and quantum metrology. Ideal PNR detection remains experimentally challenging, motivating both intrinsic energy-resolving detectors and multiplexed architectures based on simpler click detectors [1,2].

Transition-edge sensors provide high-fidelity photon-number resolution by measuring absorbed energy, but their thermal response can limit the achievable repetition rate, which affects their applicability [3,4]. On the contrary, multiplexed click detectors and superconducting nanowire single photon-detector (SNSPD) arrays allow faster timing performance but provide only approximate PNR [5]. In fact, a single SNSPD has traditionally been regarded as capable of distinguishing only between the absence and the presence of one or more photons, without providing information on photon-number or energy. Therefore, it is operated as binary on/off detector.

Spatial multiplexing uses arrays of SNSPDs with independent readouts architecture but can also be implemented with nanowires arranged in parallel [6,7]. However, multiplexing techniques yield only quasi-PNR capabilities, because each resolving element remains binary. Therefore, fidelity is limited by multi-photon collisions within the same pixel, and the number of multiplexed elements needed to keep this collision probability low scales quadratically with the photon-number $N$ [2], often leading to highly complex system architectures [8-10]. It is natural to ask whether this requirement can be relaxed if each individual element already carries some photon-number information of its own.

Intrinsic PNR in a single SNSPD was first demonstrated by Cahall et al. [11, 12], who showed that the electrical traces of SNSPD contain photon-number-dependent features. Recent works have shown that arrival-, rise- and fall-time statistics can carry information about multi-photon absorption events that can be easily accessed using moder ultra high time-resolution time taggers [13-16]. In particular, high-fidelity discrimination of zero-, one- and two-or-more-photon events has recently been demonstrated in a commercial SNSPD under high-repetition-rate pulsed illumination, with a conditional two-photon misidentification probability equivalent to that of a multiplexed array of approximately 730 independent pixels [17], which a clear example of how much information a single, individually resolving element can already provide.

Explicit formulas for photon-number discrimination with on/off detectors have been worked out in [18] allowing to compute the probability of observing $C$ clicks by distributing a Fock state of $N$ photons evenly among $D$ on/off (2-state) detectors with quantum efficiency $\eta$ belonging to a multiplexed system. In this work, the authors also present a Bayesian approach to retrodict the photon-number given an observed number of clicks.

This line of work has since been extended to compare multiplexing architectures through information-theoretic figures of merit [19] and refined to bound the achievable estimation error [1], and generalized to waveguide-integrated array geometries [20], but in all cases the local detector model remains strictly binary. Separately, spatial multiplexing has been combined with per-element intrinsic resolution, either via a different readout mechanism [21] or for a fixed two-photon regime [22]. However, these formulas are not applicable to, and no systematic combinatorial and Bayesian framework yet exists for, multiplexed systems built from detectors that are each individually capable of resolving more than two states — a direction explicitly identified as a promising route to extend photon-number resolution beyond current limits [23], and one which has attracted our attention.
These developments motivate the central goal of this work, that is quantifying how much additional information is gained, and whether the quadratic scaling with $N$ discussed above can be relaxed, when the local detector model is refined from on/off to multi-state photon-number resolution. This is done, not only by computing the response probability of an array of multi-state detectors— thereby generalizing the forward formulas of [18] —, but also by inferring the statistics of photon-number state of the source, given the response of the detectors, generalizing the corresponding Bayesian retrodiction. We develop a combinatorial and Bayesian framework for a generic multiplexed detection system consisting of a number of detectors, each capable of resolving an arbitrary number of photons. We account for the effects of limited detector efficiency and propose a formulation that is further extended, in the Appendix, to include the dark counts noise. However, since the typical sub-hertz rate of dark counts in SNSPDs provides negligible effects on fidelity compared to the collision-driven and resolution-driven contributions [17], we show numerical results of our models for noiseless 4- and 8-detector multiplexed systems based on 2-, 3-, and 4-state detectors, individually capable of distinguishing 0- and 1plus-photon, 0-1-and 2plus-photon, 0-1-2- and 3plus-photon events, respectively.

*Contact author: fabio.auriemma@spin.cnr.it

## II. GENERALIZED MULTIPLEXED DETECTION MODEL

### *A. Probability response tensor*

A multiplexed system composed of an array of $D$ detectors (nanowires) provides different response depending on the PNR capability of the constituting detectors. An s-state detector provides s output classes, corresponding to 0,1,…,$s$−2 resolved photons and a final saturated class, ($s$−1)+, which groups all occupancies equal to or larger than (s−1) photons. For a single detection event triggered by the arrival of N photons on the multiplexed system, we define the state (or output) vector, $\boldsymbol{K}^{(s)} = (k_0, k_1, \dots, k_{s-1})$, whose components represent the number of detectors that sense 0,1, 2, …($s$-1)+ photons. Obviously, the sum of the components of the state vector, is $\sum_{i=0}^{s-1} k_i = D$. By defining the counting vector $\boldsymbol{v}^{(s)}$ =(0,1,…$s$-1), the photon-number readout is given by the scalar observable (or estimator) $N_s = \boldsymbol{K}^{(s)} \cdot \boldsymbol{v}^{(s)}$. Due to possible saturation of one or more detectors, $N_s$ represents the lower-bound estimator associated with the $s$-state detector, i.e. generally $N_s(\boldsymbol{K}) \leq N$. For instance, in a 3-state device (s=3), each detector can discriminate 0,1 and 2+ photons, and a system constituted by $D$ detectors of this type will output the state vector $\boldsymbol{K}^{(3)} = (k_0, k_1, k_2)$ with $k_0 + k_1 + k_2 = D$. In this case $\boldsymbol{v}^{(3)} = (0,1,2)$ and the maximum photon-number that can be counted will be $\boldsymbol{K}^{(3)} \cdot \boldsymbol{v}^{(3)} = 0k_0 + 1k_1 + 2k_2$. We will often particularize our model to three types of states, using $s$=2,3 and 4, representing spatially multiplexed systems constituted by 2-, 3- and 4-state detectors, respectively.
For the sake of notational clarity, we will drop the superscript ($s$) to vectors $\boldsymbol{K}^{(s)}$ and $\boldsymbol{v}^{(s)}$ any time it is not strictly necessary to highlight the state number s of the detector or the multiplexed system.
Mapping the probability of a certain global output $\boldsymbol{K}^{(s)}$ given $N$, for different s and a fixed $D$, i.e. finding P($\boldsymbol{K}^{(s)}$|$N,D$), is the first task of this section. In its most general form, the detector response is a probability response tensor since the outcome $\boldsymbol{K}^{(s)}$ is a multi-component vector defined by the occupancy classes of the detectors in a specific detection event. When this vector outcome is mapped onto the scalar observable $N_s$, the tensor is marginalized into a two-dimensional matrix $P(N_s \mid N)$.
If the same detection event would be sensed by two multiplexed systems with the same number of detectors but different sensing capabilities, i.e. $s$=$\alpha$ for the first system and $s$= $\alpha$+$\beta$ for the second one, hierarchical relationships can be affirmed between the components of the output vector $\boldsymbol{K}^{(\alpha)}$ and $\boldsymbol{K}^{(\alpha+\beta)}$, $k_i^{(\alpha)}$ *(i=0, … α-1)* and $k_j^{(\alpha+\beta)}$ *(j=0, … , α+β-1)*:

$$k_i^{(\alpha)} = k_j^{(\alpha+\beta)}, \quad i = j \leq \alpha - 2 \tag{1}$$

and

$$k_{\alpha-1}^{(\alpha)} = \sum_{j=\alpha-1}^{\alpha+\beta-1} k_j^{(\alpha+\beta)} \tag{2}$$

We clarify the meaning of these two expressions by referring to a 3-state and a 4-state system, i.e. $\boldsymbol{K}^{(3)} = (k_0^{(3)}, k_1^{(3)}, k_2^{(3)})$ and $\boldsymbol{K}^{(4)} = (k_0^{(4)}, k_1^{(4)}, k_2^{(4)}, k_3^{(4)})$, i.e. $\alpha = 3, \beta = 1$ . In this case, the 0- and 1 -photon-numbers can be unambiguously distinguished by the detectors of both multiplexed systems, while there is no certainty about 2 and more than 2 photons in case of the 3-state system. Eq.1 expresses that the number of detectors that sense 0- and 1-photon do not change if we use detectors with better resolving capabilities, e.g., $k_0^{(3)} = k_0^{(4)}$ and $k_1^{(3)} = k_1^{(4)}$. Eq.2, instead, expresses that the number of detectors which sense 2 and more than 2 photons in a 3-state system equals the sum of the number of detectors with better resolving capabilities that sense 2, 3 and more than 3 photons, i.e. $k_2^{(3)} = k_2^{(4)} + k_3^{(4)}$. Similarly, one can write, $k_2^{(3)} = k_2^{(5)} + k_3^{(5)} + k_4^{(5)}$ ($j$=5).

We use exponential generating functions to map $\boldsymbol{K}^{(s)}$, for fixed *D*, *s* and *N*. As s is now fixed, for the sake of notational clarity we will drop the superscript (s) to the components of $\boldsymbol{K}^{(s)}$. As known from combinatorics [24], the problem of finding the number of ways $k_1$ specific detectors receive 1 photon, $k_2$ detectors receive 2 photons, and so on, up to the ks-1 saturated detectors receiving s-1 plus photons, can be recast into finding the number of solutions of the following constrained equation:

$$x_1 + x_2 + \cdots + x_{k_1} + \cdots + x_{k_1+k_2} + \cdots + \cdots + x_{k_1+k_2+\cdots k_{s-1}} + \cdots + x_D = N \tag{3}$$

with :

$$x_i \begin{cases} 1\,, & i \le k_1 \\ 2\,, & k_1 < i \le k_1 + k_2 \\ \cdots & \\ s-2\,, & k_1 + \cdots k_{s-3} < i \le k_1 + \cdots k_{s-2} \\ a \in \{s-1, s, s+1, \dots\}\,, & k_1 + \cdots + k_{s-2} < \\ & < i \le k_1 + \cdots + k_{s-1} \\ 0, & k_1 + \cdots + k_{s-1} < i \le D \end{cases} \tag{4}$$

where the 1 (ones) terms correspond to $k_1$ detectors, the 2 terms to $k_2$ detectors and so on, up to the s-2 terms. The $x \in \{s-1, s, s+1, \dots\}$ terms represent possibly saturated $k_{s-1}$ detectors and the zero terms correspond to $k_0=D-k_1-k_2-\dots-k_{s-1}$ detectors which receive no photons.

Using exponential generating function, the solution to this problem will be $N!\,[x^N]f(x)$ , where the term $[x^N]f(x)$ is simply the coefficient of the monomial term $x^N$ within the exponential generated function $f(x)$:

$$f(x) = \left(\frac{x^0}{0!}\right)^{k_0} \left(\frac{x^1}{1!}\right)^{k_1} \left(\frac{x^2}{2!}\right)^{k_2} \cdots \left(\frac{x^{s-2}}{(s-2)!}\right)^{k_{s-2}} \left(\frac{x^{s-1}}{(s-1)!} + \frac{x^s}{s!} + \cdots\right)^{k_{s-1}} =$$
$$= \frac{x^{1k_1+2k_2+\cdots(s-2)k_{s-2}}}{0!^{k_0}\,1!^{k_1}\,2!^{k_2} \dots (s-2)!^{k_{s-2}}} \left(e^x - 1 - x - \frac{x^2}{2!} - \cdots - \frac{x^{s-2}}{(s-2)!}\right)^{k_{s-1}} \tag{5}$$

In the expression of *f(x)*, all the terms of the type $\left(\frac{x^i}{i!}\right)^{k_i}$ unequivocally correspond to the number of ki detectors which are able to unequivocally count i photons. Since an *s*-state detector is able to unequivocally resolve up to *s*-2 photons, *i* ranges from *0* to *s*-2. On the contrary, the indistinguishability between s-1 and more photons reflects in the terms in parenthesis powered to $k_{s-1}$. In the second equality we have used the definition of the expansion of the exponential function $e^x$. Using properties of powers, the solution to the problem can be rewritten as

$$N!\,[x^N]f(x) = N!\left[x^{N-\boldsymbol{K}^{(s)}\cdot\overline{\boldsymbol{v}}^{(s)}}\right]g(x) \tag{6}$$

In Eq. 6,

$$g(x) = \frac{1}{0!^{k_0}\, 1!^{k_1}\, 2!^{k_2} \dots (s-2)!^{k_{s-2}}} \left( e^x - 1 - x - \frac{x^2}{2!} - \dots - \frac{x^{s-2}}{(s-2)!} \right)^{k_{s-1}} \tag{7}$$

and $\overline{\boldsymbol{v}}^{(s)} = \{v^{(s)}, 0\} = \{0,1, \dots, s-1,0\}$. In other words, instead of finding the coefficient of the monomial term $x^N$ of the $f(x)$ in Eq.5, one can find the coefficient of the monomial term $x^{N-1k_1+2k_2+\cdots(s-2)k_{s-2}}$ of the $g(x)$ in Eq.7.
The final computation for the probability we are looking for, will account for any set of $\boldsymbol{K}^{(s)}$ detectors from the total number of detectors *D*. Therefore, the solution to the combinatorial problem must be multiplied by $\frac{1}{D^N}$, which is the total number of possible outputs, $by$ $\binom{D}{k_1}$ is the number of ways k1 detectors can be chosen out of D, by $\binom{D-k_1}{k_2}$ i.e. the number of ways $k_2$ detectors can be chosen out of *D* once $k_1$ have been taken, and so on, up to the last binomial coefficient accounting for the number of ways $k_s$ detectors can be chosen from the remaining ones. By using the properties of the multinomials – see, for instance Ref [25] – and knowing that $k_0=D-k_1-k_2-\dots-k_{s-1}$, one reaches the compact form for the ideal probability response tensor:

$$P\left(\boldsymbol{K}^{(s)}|N,D\right) = \frac{1}{D^N} \frac{D!}{\boldsymbol{K}^{(s)}!} \frac{N!}{\prod_{r=1}^{s-2}(r!)^{k_r}} \times \left[x^{N-\boldsymbol{K}^{(s)}\cdot\overline{\boldsymbol{v}}^{(s)}}\right] \left( e^x - \sum_{r=0}^{s-2} \frac{x^r}{r!} \right)^{k_{s-1}} \tag{8}$$

Where $\boldsymbol{K}^{(s)}! = k_0!\, k_1! \dots k_{s-1}!$. The probability response tensor obtained above is an ideal one, because so far, the detection inefficiencies of the nanowires have not been accounted for.
We can particularize the results obtained so far to practical cases:

- For 2-state detectors:
  $f^{(2)}(x) = (e^x - 1)^{k_1}$ and $N - \boldsymbol{K}^{(2)} \cdot \overline{\boldsymbol{v}}^{(2)} = N$.
- For 3-state detectors:
  $f^{(3)}(x) = x^{k_1}(e^x - 1 - x)^{k_2}$ and $N - \boldsymbol{K}^{(3)} \cdot \overline{\boldsymbol{v}}^{(3)} = N - k_1$.
- For 4-state detectors:
  $f^{(4)}(x) = x^{k_1} \left(\frac{x^2}{2!}\right)^{k_2} \left(e^x - 1 - x - \frac{x^2}{2!}\right)^{k_3}$ and $N - \boldsymbol{K}^{(4)} \cdot \overline{\boldsymbol{v}}^{(4)} = N - k_1 - 2k_2$.

Therefore:

$$P\left(\boldsymbol{K}^{(2)}|N,D\right) = \frac{1}{D^N} \binom{D}{k_1} N!\, [x^N](e^x - 1)^{k_1} \tag{9}$$

$$P\left(\boldsymbol{K}^{(3)}|N,D\right) = \frac{1}{D^N} \frac{D!}{k_1!\, k_2!\, (D - k_1 - k_2)!} N!\, [x^{N-k1}](e^x - 1 - x)^{k_2} \tag{10}$$

$$P\left(\boldsymbol{K}^{(4)}|N,D\right) = \frac{1}{D^N} \frac{D!}{k_1!\, k_2!\, k_3!\, (D - k_1 - k_2 - k_3)!} \frac{N!}{(2!)^{k_2}} [x^{N-k_1-2k_2}] \left( e^x - 1 - x - \frac{x^2}{2!} \right)^{k_3} \tag{11}$$

For *s*=2, Eq. 9 reduces exactly to the Lemma of [18] expressing the probability of observing *C* clicks by distributing a Fock state of *N* photons evenly among *D* ideal on/off detectors.

### *B. Lossy probability response tensor*

Real detectors are provided with efficiency $\eta<1$, thus the number of photons effectively registered by the detector array is $R \leq N$. The probability that exactly *R* photons will survive the absorption process follows a global binomial distribution. Since the spatial splitting of photons across the array is perfectly uniform, a lossy system with *N* incident photons behaves identically to an ideal system injected with a reduced number of photons effectively detected, *R*. Therefore, the realistic, lossy probability response tensor is computed by performing a Binomial Loss Convolution – see Ref [26] for instance – over the ideal states:

$$P\big(\boldsymbol{K}^{(s)}|N,D,\eta\big)=\sum_{R=\boldsymbol{K}^{(s)}\cdot\boldsymbol{\nu}^{(s)}}^{N}\binom{N}{R}\eta^{R}(1-\eta)^{N-R}\,P_{ideal}\big(\boldsymbol{K}^{(s)}|\,R,D\big) \tag{12}$$

The dependence of $P$ on $\eta$ in Eq.12 implies that the system is lossy, as opposed to the $P$ in Eq.1 that refers to an ideal lossless system. The four terms in the product within the summation in Eq.12 represent, from left to right, the number of ways $R$ out of $N$ photons can be chosen, the probability the $R$ photons are detected by devices with efficiency $\eta$, the probability that $N$-$R$ photons are not detected by these devices (in other words, the first three terms of the product represent $P(R|N,\eta)$ ) and, finally, the ideal probability tensor computed for R incident photons in an s-state multiplexed system. The lower bound of the summation represents the minimum number of photons that can trigger the response $\boldsymbol{K}^{(s)}$. For instance, for a 3-state detector, $R_{min}= 1k_1 + 2k_2$, since it is physically impossible to trigger $k_1$ single-clicks and $k_2$ overflow-clicks with fewer than $1k_1 + 2k_2$ absorbed photons. Eq.12 generalizes the noiseless detection corollary of Ref [18] accounting for multiplexed noiseless on/off detectors with quantum efficiency η.
Consistency relations between the detector models can be carried out by starting from the hierarchical relationships, Eq.2, which translate into probabilistic relationships where the $s$-th state can be written as a marginalization of the $(s+1)$-th state:

$$P\big(\boldsymbol{K}^{(s)}|N,D,\eta\big)=\sum_{k_{s-1}^{(s)}=k_{s-1}^{(s+1)}+k_{s}^{(s+1)}}P\big(\boldsymbol{K}^{(s+1)}|N,D,\eta\big) \tag{13}$$

Where we have used the superscripts ($s$) and ($s$+1) to distinguish the components of the $\boldsymbol{K}^{(s)}$ vectors of the ($s$) state device from the vectors $\boldsymbol{K}^{(s+1)}$ of the ($s$+1)-state device. For instance, the 3-state tensor is a marginalization of the 4-state one, in fact taking $s$=4 yields and writing the state vectors in terms of their components yields

$$P^{(3)}\big(k_0^{(3)}k_1^{(3)},k_2^{(3)}|N,\eta\big)=\sum_{k_2^{(3)}=k_2^{(4)}+k_3^{(4)}}P^{(4)}\big(k_0^{(4)},k_1^{(4)},k_2^{(4)},k_3^{(4)}|N,\eta\big) \tag{14}$$

In Eq. 14, $k_0^{(3)}=k_0^{(4)}$ and $k_2^{(3)}=k_2^{(4)}$ due to Eq.1. Similarly, the 2-state detector is a marginalization of the 3-state detectors by constraining the summation of the probabilities with $k_1^{(3)}+k_2^{(3)}=k_1^{(2)}$.

### *C. Photon-number-state reconstruction*

To evaluate the true performance of a multiplexed system comprising an array of photon-number-resolving detectors, the probability tensor $P\big(\boldsymbol{K}^{(s)}\mid N,D,\eta\big)$, derived in Sec. II.b., must be inverted. In other words, we wish to infer the likelihood that N photons were incident given the observation of a specific detection vector $\boldsymbol{K}^{(s)}$, which is referred to as the posterior probability distribution. This inversion is handled rigorously using Bayes' Theorem [27]. To this aim, an assumption about the prior distribution, i.e. the quantum statistical distribution of the light emitted by the source, is required, $P_{prior}(N\mid\boldsymbol{\Theta})$, where $\boldsymbol{\Theta}$ indicates the generic parameters of the input state. For instance, $P_{prior}(N|\boldsymbol{\Theta})$ may be a poissonian distribution with mean photon-number $\mu$. In this case:

$$P_{prior}(N|\mu)=\frac{e^{-\mu}\mu^{N}}{N!} \tag{15}$$

Applying Bayesian inference, the posterior probability distribution is reconstructed by combining the detector response model with the prior photon statistics of the light source, as

$$P_{Bayes}\big(N\mid\boldsymbol{K}^{(s)},D,\eta,\boldsymbol{\Theta}\big)=\frac{P\big(\boldsymbol{K}^{(s)}\mid N,D,\eta\big)\,P_{prior}(N\mid\boldsymbol{\Theta})}{\sum_{N'=N_s}^{N_{max}}P(\boldsymbol{K}^{(s)}\mid N',D,\eta)\,P_{prior}(N'\mid\boldsymbol{\Theta})} \tag{16}$$

where the denominator represents the global evidence $P(K)$, which acts as a statistical normalization factor. In principle, the normalization sum should run from $N_{min}$=0 to $N_{max}$=∞, since any non-negative number of incident photons is allowed by the prior distribution. In numerical implementations, this infinite sum is truncated at a finite cutoff $N_{max}$, implying the presence of a residual error that can be estimated and controlled. Since probabilities cannot exceed unity, the error of the prior distribution is larger than the total residual error at the denominator:

$$R_{\boldsymbol{k}}(N_{max}) = \quad \sum_{N\prime=Nmax+1}^{\infty} P\big(\boldsymbol{K}^{(s)} \mid N', D, \eta\big)\, P_{prior}(N' \mid \boldsymbol{\Theta}) \tag{17}$$

Therefore, by controlling the error on the prior distribution we limit the total residual error. A lower bound is used too, $N_{min} = N_s = \boldsymbol{K}^{(s)} \cdot \boldsymbol{v}^{(s)}$ because the likelihood vanishes for photon-numbers smaller than that compatible with the vector $\boldsymbol{K}^{(s)}$ when dark counts are neglected.

## III. FIDELITIES

### *A. N-photon detection fidelity*

The $N$-photon detection fidelity (or simply *fidelity*) is the probability that an input pulse containing $N$ photons is reconstructed as exactly $N$ photons by a given detection device.
Since an $s$-state detector resolves exactly the occupations $0,1,\dots,s-2$, while the last class corresponds to $(s-1)+$ photons, the count provided by the multiplexed system, $N_s$, is a lower-bound estimate of the true photon-number. Therefore, given N incident photons, an output satisfying $N_s = N$ corresponds to an exact reconstruction, and it occurs if none of the detectors in the saturated $(s-1)+$ class contains more than $s-1$ photons. Saturation ambiguity can only arise in the $k_{s-1}$detectors belonging to this last class, since all lower occupations are resolved unambiguously.
In principle, one could estimate the detector fidelity of the s-state system by removing the probabilities of saturation from the probabilities that the system reads $N_s = N$. However, a more direct way to compute it is to consider 2 systems, with the same $D$, $\eta$ and incident photons $N$, but one provided with s-state detectors and another one with ($s$+1)-state detectors. To compute the fidelity of the first system, we assume $N_s = N$ and use the probability tensor of the ($s$+1)-state system by imposing:

$$\boldsymbol{K}^{(\boldsymbol{s}+\boldsymbol{1})} = (\boldsymbol{K}^{(s)}, 0) \tag{18}$$

In fact, conditioning $N_s = N$ implies that all detectors of the $s$-state system remain non-saturated, and possible $k_{s-1}^{(s+1)}$ detectors receive exactly $s$-1 photons. The same detection is realized with the ($s$+1)-state system by imposing $k_s^{(s+1)}$= 0, which is precisely the condition imposed by Eq.18, i.e. $k_0^{(s+1)} = k_0^{(s)}, \dots, k_{s-1}^{(s+1)} = k_{s-1}^{(s)}, k_{s-1}^{(s+1)} = 0$, representing the only extension of an s-state output vector to an (s+1)-state output vector occurring with non-saturated detectors. Therefore, the $N$-photon detection fidelity of the first system can be computed as:

$$F^{(s)}(N) = \sum_{\boldsymbol{K}^{(s)}\cdot\boldsymbol{v}^{(s)}=N} P^{(s+1)}\big((\boldsymbol{K}^{(s)}, 0) | N, D, \eta\big) \tag{19}$$

The conditional summation accounts for probabilities that result in the same count $N$ for different output configurations $\boldsymbol{K}^{(s)}$. For instance, a 8-detector system made of a 3-state detectors could output $\boldsymbol{K}^{(\boldsymbol{3})} = (k_0, k_1, k_2) = (3,2,3)$ counting $\big(\boldsymbol{K}^{(3)} \cdot \boldsymbol{v}^{(3)}\big) = 0 * 3 + 1 * 2 + 2 * 3 = 8$ photons. However, one of the k2 detectors may saturate, if it receives more than 2 photons, and the actual number of incident photons could be larger than 8. This is why we use the probability for the 4-state detectors imposing (3,2,3,0) output vector. Moreover, the same count of 8 photons can be provided also by other configurations, e.g. (1,6,1), (2,4,2), … and we use the conditional summation because each of them has a certain probability of occurring.

### *B. Reconstruction fidelities*

The detector fidelity characterizes the detector as a counting device, because it provides an operational measure of the exact-counting capability of the multiplexed detection system. However, in many applications it could be useful to characterize the ability of the detector to reconstruct the actual incident photonic state, given the photon-number distribution of the light source (e.g. poissonian, $\mu$). Therefore, we introduce an additional fidelity, the reconstruction fidelity which characterizes the detector as an inference device, by measuring the probability that the posterior photon-number distribution, $P_{Bayes}(N \mid \boldsymbol{K}^{(s)}, D, \eta, \mu)$, assigns to the true photon-number. In this way we can assess the capability of transforming an output $\boldsymbol{K}^{(s)}$ into a reliable distribution of photons.

We will see from the results that saturated outcomes affect the two fidelities in different ways. In the detector fidelity, a saturated class contributes to failure only when it causes the readout $N_s$ to underestimate the true photon-number. For example, in a 4-state detector, the 3 class is still compatible with exact reconstruction when $N = 3$. In Bayesian reconstruction, however, the same saturated outcome remains ambiguous because it can also be generated by higher photon-numbers, and therefore the posterior probability assigned to the true $N$ is reduced.

We use Bhattacharyya coefficient for the fidelity of the reconstructed state of photon-number detectors, as also done to validate reconstructed vs. simulated POVMs in [16]. It provides classical fidelity of the reconstructed state, which is ultimately our goal. In fact, since the detector response is photon-number based, it provides the projection of the quantum state on Fock basis. Therefore, the phase coherence among Fock states is lost and we reconstruct a classical distribution of number of photons with a randomized phase, not the entire matrix of the optical state. Therefore, in our case, the Uhlmann quantum fidelity reduces to the Bhattacharyya (classical) fidelity [28, 29].

We must recall that in quantum optics, given a true photon distribution, $P_{true}(N)$, and a reconstructed (estimated) one, $P_{est}(N)$, the Bhattacharyya coefficient is used to measure their overlap or statistical similarity, and it is a measurement of the classical fidelity. It is defined as:

$$F_B = \left[\sum_{N=0}^{N_{max}} \sqrt{P_{est}(N) P_{true}(N)}\right]^2 \tag{20}$$

If the distributions match perfectly, the fidelity becomes 1, if they overlap partially this fidelity reduces. For instance, for a single 4-state detector and 4 incident photons, the only contribution to the summation in Eq.20 will come from $N$=4, and it will be less than one because $P_{true}(4) = 1$ but $P_{est}(4|4) < 1$. In fact, the output 4 can be produced in the 4-state detector by 4, 5, or more incident photons.

In case of photon detectors, the Bhattacharyya coefficient provides the local reconstruction fidelity, the estimated guess on the input state, given by the Bayesian inference, is $P_{est}(N) = P_{Bayes}(N \mid \boldsymbol{K}^{(s)}, D, \eta, \boldsymbol{\Theta})$, for a specific output vector $\boldsymbol{K}^{(s)}$, while the true input state is the probability distribution of the specific light pulse, $P_{true}(N) = \delta_{N,N_{true}}$, i.e. the Kronecker delta centered on the true photon-number of that specific impulse, as seen in the example above. The local reconstruction fidelity of a single output $\boldsymbol{K}^{(s)}$ of the detection system is:

$$\mathcal{F}_{l.r.}(\boldsymbol{K}^{(s)}, D, \eta, \boldsymbol{\Theta}, N_{true}) = \left[\sum_{N=0}^{N_{max}} \sqrt{P_{Bayes}(N \mid \boldsymbol{K}^{(s)}, D, \eta, \boldsymbol{\Theta}) \delta_{N,N_{true}}}\right]^2 \tag{21}$$

Which means:

$$\mathcal{F}_{l.r.}(\boldsymbol{K}^{(s)}, D, \eta, \boldsymbol{\Theta}, N_{true}) = P_{Bayes}(N_{true} \mid \boldsymbol{K}^{(s)}, D, \eta, \boldsymbol{\Theta}) \tag{22}$$

In this definition, the prior complete photon distribution of the source only appears in the Bayesian inference. Using $P_{true}(N) = P_{prior}(N, \boldsymbol{\Theta})$ would compare the specific detection output with the entire prior distribution, providing an inconsistent fidelity that favors less-informative systems.

Obviously, to evaluate the global performance of the detection system, looking at just one specific output $\boldsymbol{K}^{(s)}$ is not sufficient. We obtain the global fidelity of the reconstructed state as the statistical expectation value of the local fidelity across all possible patterns that the detector could show. This is given by the weighted average of the local fidelity across all possible outputs $\boldsymbol{K}^{(s)}$, where the weights are given by the probability that the specific $\boldsymbol{K}^{(s)}$ occurs.

$$\mathcal{F}_{g.r.}(D, \eta, \boldsymbol{\Theta}, N_{true}) = \mathbb{E}\left[\mathcal{F}_{l.r.}(\boldsymbol{K}^{(s)})\right] = \sum_{\boldsymbol{K}^{(s)}} P(\boldsymbol{K}^{(s)} \mid N_{true}, D, \eta) \mathcal{F}_{l.r.}(\boldsymbol{K}^{(s)}, D, \eta, \boldsymbol{\Theta}, N_{true}) \tag{23}$$

Using the result from Eq.22, the global fidelity of the reconstructed state becomes:

$$\mathcal{F}_{g.r.}(D, \eta, \boldsymbol{\Theta}, N_{true}) = \sum_{\boldsymbol{K}^{(s)}} P(\boldsymbol{K}^{(s)} \mid N_{true}, D, \eta) P_{Bayes}(N_{true} \mid \boldsymbol{K}^{(s)}, D, \eta, \boldsymbol{\Theta}) \tag{24}$$

In this sense, $\mathcal{F}_{g.r.}$ "scans" over all possible output states $\boldsymbol{K}^{(s)}$ and evaluates the functioning of the entire chain source-detector-inference associated with detection of a single photon-number, $N_{true}$.
The reconstruction fidelity is a stringent criterion for two main reasons: 1. This fidelity depends not only on whether the detector readout equals the true photon-number, but also on how much posterior probability is assigned to this number after observing a specific outcome $\boldsymbol{K}^{(s)}$; 2. The true photon-number accounted for may have itself a low probability to occur for a specific prior distribution. If, for instance, a 4-state system provided with $D$=4, the true photon-number $N_{true}$=10 could produce an outcome like $\boldsymbol{K}^{(4)}$= (0,1,1,2). However, the same outcome may derive from 11, 12, etc., if more than 3 photons arrive on the saturated detectors. Therefore, the posterior reconstruction will never unequivocally identify the true photon-number but will remain distributed over several photon-numbers close to the true one. Furthermore, if for instance the light source is a poissonian with $\mu$=6, the prior probability will deteriorate the fidelity as the probability for the source to produce $N_{true}$=10 photons is reduced.
The final step is evaluating the reconstruction performance over an ensemble of pulses, instead of a single one. To this aim, we average the global fidelity of the reconstructed state over the statistical distribution of the source, $P_{source}(N|\boldsymbol{\Lambda})$, $\boldsymbol{\Lambda}$ denoting the parameters of the source state (e.g. the mean photon-number $\mu$ in case of a poissonian source). For now, we consider $P_{source}(N|\boldsymbol{\Lambda})$ coinciding with the prior distribution, $P_{prior}(N\,|\boldsymbol{\Theta})$ appearing in the Bayesian reconstruction, Eq.16. Since the global fidelity itself has been defined as an average over the output states for a specific true photon-number, the ensemble reconstruction fidelity is then obtained by averaging the local fidelity over both the detector outcomes $\boldsymbol{K}^{(s)}$ and the prior distribution:

$$F_{ensemble}(D,\eta;\boldsymbol{\Theta}) = \sum_{N=0}^{N_{max}} P_{prior}(N|\boldsymbol{\Theta}) \sum_{\boldsymbol{K}^{(s)}} P\left(\boldsymbol{K}^{(s)} \mid N, D, \eta\right) P_{Bayes}\left(N \mid \boldsymbol{K}^{(s)}, D, \eta, \boldsymbol{\Theta}\right) \tag{25}$$

In this expression $P_{Bayes}$ still represents the whole local fidelity term, as in Eq.22. This means that the $P_{true}(N)$ utilized when applying Eq.22 for the local fidelity, is still a Kronecker delta, not the source distribution $P_{prior}(N|\boldsymbol{\Theta})$. In fact, also in this case, the local fidelity is defined for single individual pulses.
The prior distribution appears, in Eq.25, implicitly in the $P_{Bayes}$ term (according to Eq.16), and explicitly as the weighting terms representing the photon source statistics.
However, in order to quantify the robustness of the reconstruction against prior mismatch or against deviations of the actual source statistics, one could further generalize Eq.25 by replacing the explicit term with another – generic – distribution, $P_{source}(N|\boldsymbol{\Lambda})$.

## IV. RESULTS AND DISCUSSION

To show the impact that different state detectors have on the overall performance of a multiplexed system we consider 2-, 3- and 4-state detectors. Once $D$, and $\eta$ are fixed, we can project the full probability tensor $P^{(s)}\left(\boldsymbol{K}^{(s)}|N,D,\eta\right)$ onto the scalar estimator $N_s = \boldsymbol{K}^{(s)} \cdot \boldsymbol{v}^{(s)}$ at different $N$. This yields the likelihood matrix $P^{(s)}(n_s|N,D,\eta)$, which is obtained by marginalizing the full s-state response tensor over all output configurations $\boldsymbol{K}^{(s)}$ that yield the same estimator value $n_s$ :

$$P^{(s)}(n_s|N,D,\eta) = \sum_{\boldsymbol{K}^{(s)}:\, N_s=n_s} P^{(s)}_{lossy}\left(\boldsymbol{K}^{(s)}|N,D,\eta\right) \tag{26}$$

Therefore, the likelihood matrix of the 2-state systems coincides with the detector tensor, $P^{(2)}(n_2|N,D,\eta) = P^{(2)}_{lossy}(k_1|N,D,\eta)$, while $P^{(3)}(n_3|N,D,\eta) = \sum_{1k_1+2k_2=n_3} P^{(3)}_{lossy}(k_1,k_2|N,D,\eta)$ for the 3-state system, and $P^{(4)}(n_4|N,D,\eta) = \sum_{1k_1+2k_2+3k_3=n_4} P^{(4)}_{lossy}(k_1,k_2,k_3|N,D,\eta)$ for the 4-state system. Through the likelihood matrix way we can directly compare the photon-number readout of different state detectors when $N$ varies.
Figure 1 shows the likelihood matrices for $D$=8, $\eta$=0.92 when $N$ is varied up to 16. The white diagonal shows an ideal response, where the number of arriving photons equals the readout output. The heatmaps show that, when the local photon-number resolution increases, the detector response becomes progressively closer to the ideal diagonal.
In fact, the response of the 2-state system is strongly compressed because, as $N$ increases, the number of detected photons increases with a reduced rate.

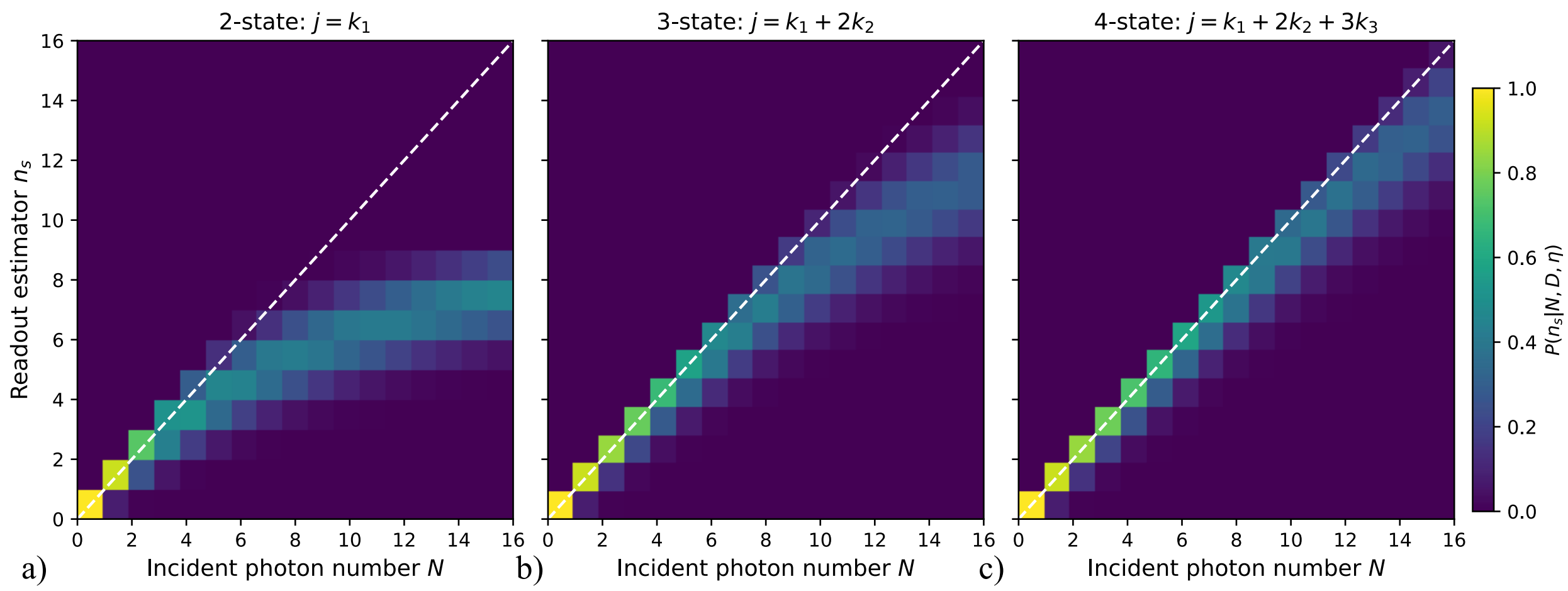

Figure 1. Projected likelihood matrices $P^{(s)}(n_s|N, D, \eta)$ for $D$=8 and $\eta$=0.92. a) 2-state system; b) 3-state system; c) 4-state system.

The 3-state system shows a likelihood that better distributes along values of $n_3$ that are closer to N. Finally, the 4-state likelihood matrix is clearly more concentrated along the diagonal, as collisions and losses remain limited. Therefore, the projected readout is more informative with respect to the number of incident photons when the state number increases. The probability tensor finds immediate application also for the computation of the fidelity of the detectors.

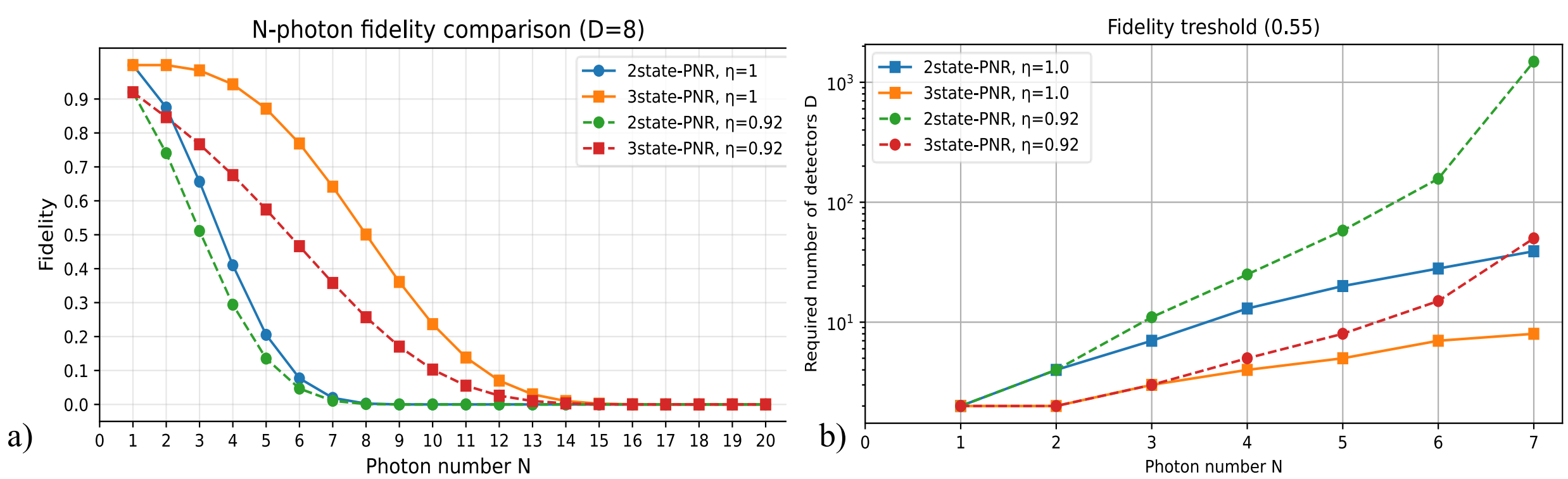

Figure 2. Detector fidelity of 2- and 3-state detection systems with $\eta$ =1 and $\eta$ =0.92. a) N-photon fidelity expressed as a function of $N$ using $D$=8; b) Number of detectors required to guarantee fidelity of 0.55 as a function of $N$.

By numerically implementing the formulation provided in Section III, we can obtain the fidelity of a 2-state system by using the probability tensor of the analogue 3-state system by imposing $k_3^{(3)} = 0$ and that of the 3-state system from the probability tensor of the analogue 4-state system when $k_4^{(4)} = 0$.

The results are shown in Figure 2. Specifically, Figure 2a shows the fidelity as a function of the photon-number for different efficiencies of the detectors. The 3-state system can provide fidelities sensibly higher than the 2-state system up to $N$=12 and the lossless 2-state system is superior to the lossy 3-state one only up to 2 photons. Figure 2b shows the number of detectors required by a multiplexed system to satisfy a minimum fidelity threshold of 0.55. The increase in number for 2-state systems with limited efficiency is growing faster than $e^N$ and it is unpractical for 7 photons (the required $D$ is greater than 1000). Instead, up to 5 photons, the requirement for 3-state systems remains below ten detectors, with minor influence of losses when $\eta \geq 0.92$, relaxing the general rule proposed in [2] for on/off detectors that high fidelity requires the number of resolving features to exceed $N^2$.

To simulate the Bayesian photon-number state reconstruction, it is necessary to provide the output vector $\boldsymbol{K}^{(s)}$ resulting from the detection of a light impulse. Once the prior distribution is known, a possible way to compare the photon state reconstructions of systems with different state-detectors could be assigning a reasonable output vector for the system with the greatest state number and retrieve, from this, the output vectors of the systems with reduced state number of detectors by using the hierarchical relationships introduced above. For instance, if the source distribution is poissonian with $\mu$=8 and $D$=6, a possible output vector for a system with 4-state detectors could be $\boldsymbol{K}^{(4)} = [2,1,2,1]$. From this vector, one obtains the corresponding state vector of the analogue system with 3-state detections, $\boldsymbol{K}^{(3)} = [2,1,3]$, and the corresponding output vector of a system with 2-state detectors, $\boldsymbol{K}^{(2)} = [2,4]$. These output vectors are sufficient to provide a fair comparison between the Bayesian reconstructions obtained from systems for a fixed output state of systems with different state numbers.

However, to obtain results which are more useful in practical applications, we have followed another approach where, instead of arbitrarily choosing a possible output vector for a single impulse, we have assigned the number of incoming photons of the impulse, $N$, and distributed them randomly among the $D$ detectors. Obviously, $N$ must be compatible with the prior distribution. The multinomial distribution provides the probability of specific photon-numbers on each detector:

$$P(n_1, n_2, \dots, n_D | N, D) = \frac{N!}{n_1!\, n_2!, \dots, n_D!} p_1^{n_1} \dots p_D^{n_D} \tag{27}$$

The terms $(n_1, n_2, \dots, n_D)$ are the components of the photon-vector of the multiplexed system, representing the number of photons received by each detector, while $p_1, \dots, p_D$ are the probabilities of each detector receiving a photon, each raised to the power of the number of photons it received. For equiprobable detectors, $p_1 = p_2 = \dots = p_D = 1/D$, thus $p_1^{n_1} \cdot p_2^{n_2} \cdot \dots \cdot p_D^{n_D} = 1/D^N$.

A numerical solver can be used to sample the stochastic vector $(n_1, n_2, \dots, n_D)$ that follows the multinomial in Eq.27, providing $N$, $D$, $p_i$ and a seed number for the random generation. From this sequence it is trivial to retrieve the state vector $\boldsymbol{K}^{(s)}$. In practice, we start with the system with the greatest state number, $s$=4 with a given $N$, obtain the sequence $(n_1, n_2, \dots, n_4)$ and, from this, the state vector $\boldsymbol{K}^{(4)}$. Following, we obtain $\boldsymbol{K}^{(3)}$ and $\boldsymbol{K}^{(2)}$, for the 3- and 2- state analogue detection systems, which are hierarchically consistent. In this way we can obtain a plausible reconstruction, which is statistically correct and useful to compare performance of detectors. However, another step is required to produce an effective microscopic reconstruction of the single impulses from a given N: one must account also for the loss in photon-numbers due to limited efficiency. To this aim, we simply call R the number of photons that are effectively detected, as opposed to $N$ that
accounts for all photons in the light impulse ($R<N$) and compute the $\boldsymbol{K}^{(s)}$ vector by using a lossy photon-vector of the multiplexed system, $(m_1, m_2, \dots, m_D)$, derived with the same procedure described above, but using $R$ instead of $N$. Mathematically, the lossy vector is obtained by applying the binomial convolution – which accounts for detection inefficiencies, as in Eq. 12 – to the probability formula, Eq.27. Therefore, the probability of the lossy photon-vector of the multiplexed system will be:

$$P(m_1 \dots m_D | N, D, \eta) = \sum_{\substack{n_1 + .. + n_D = N \\ n_i \geq m_i}} \frac{N!}{n_1!\, n_2!, \dots, n_D!} \left(\frac{1}{D}\right)^N \prod_{i=1}^{D} \binom{n_i}{m_i} \eta^{m_i} (1-\eta)^{n_i - m_i} \tag{28}$$

As for multinomial distributions, a numerical solver can be used to obtain also the stochastic vector $(m, m_2, \dots, m_D)$ from the binomial distributions applied to the multinomial ones, thus satisfying Eq.28. As before, from a possible $(m_1, m_2, \dots, m_4)$ obtained with this procedure, the state vector $\boldsymbol{K}^{(4)}$is obtained, as well as $\boldsymbol{K}^{(3)}$ and $\boldsymbol{K}^{(2)}$ with hierarchical consistency. Finally the Bayesian reconstruction can be implemented by using Eq. 16 with these state vectors and a given prior distribution. It is worth noticing that, although inefficiency is explicitly accounted for by the vector $(m_1, m_2, \dots, m_D)$, the Bayesian reconstruction must still be performed using the lossy detector response, $P(\boldsymbol{K}^{(s)} | N, D, \eta)$ within Eq.16, since the goal is to infer the original photon-number $N$, not the intermediate number of detected photons. Therefore, no double counting of losses occurs: the same efficiency parameter $\eta$ appears once in the generative simulation described by Eq.28 and once in the likelihood model used for inference, consistently describing the same physical process.

The Figure 3 is obtained by simulating detection systems with $D$=4 and $\eta$=0.92, and a poissonian light source with $\mu$ =8. By imposing $N$=8, the photon incidence pattern (3,2,2,1) was randomly returned from the multinomial distribution. Due to the limited efficiency of the detectors, the binomial convolution transformed the photon incident pattern into an effective incidence pattern (3,2,1,1), simulating the loss of one of the eight photons, finally providing $R$=7. Consequently, the output vectors for the systems with different states are: $\boldsymbol{K}^{(4)} = (0,2,1,1)$, $\boldsymbol{K}^{(3)} = (0,2,2)$ and $\boldsymbol{K}^{(2)} = (0,4)$. The Bayesian reconstruction in Fig.3 clearly shows the progressive gain in information associated with increasing detector resolution. The 2-state detector only constrains the posterior through the number of active channels, resulting in a broad distribution. The 3-state detector suppresses all photon-numbers below $N$=6, reflecting the minimum photon-numbers required by the output vector. The 4-state detector further sharpens the posterior by explicitly resolving the 2-photon and 3+-photon sectors, leading to the most concentrated reconstruction around the physically relevant region $N\approx8$, close to the prior mean $\mu=8$. Figure 3b shows that the variance of the posterior distributions decreases as the detector photon-number resolution increases as a natural consequence of Bayesian conditioning. In fact, although the prior distribution describes the unconstrained photon statistics of the source, the posterior is conditioned on the specific detector outcome. Therefore, a more informative detector response yields a more selective likelihood, thus a narrower posterior distribution.

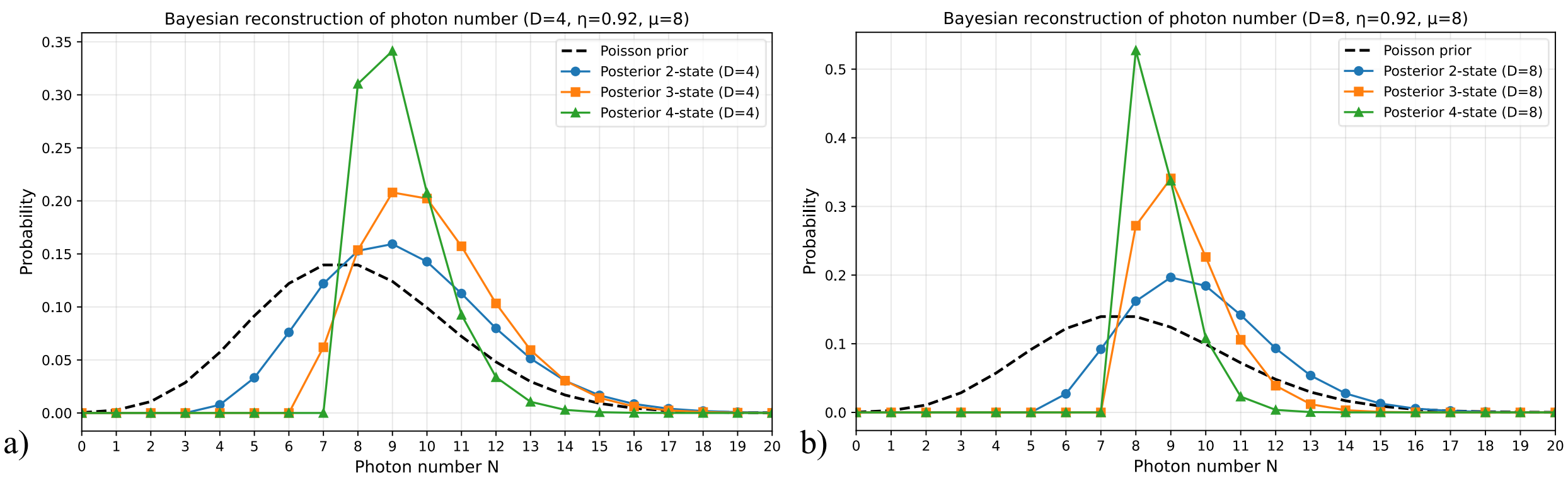

Figure 3. Bayesian reconstruction of photon-number, given prior poissonian with $\mu=8$, in detection systems with $\eta=0.92$ and a) $D=4$; b) $D=8$.

In Figure 4, the reconstruction fidelities are computed for a range of true photon-numbers, varying from 1 to 15, for systems with $D=4$ and $D=8$, and 2-,3-, and 4-state detectors. We have used a prior poissonian distribution with $\mu=8$ in Fig. 4a and with $\mu=4$ in Fig. 4b. In all cases the reconstruction fidelity remains relatively low because, as anticipated in Section III, photon-numbers are not unequivocally attributed through the Bayesian posterior to a state vector which can contain saturated elements. Obviously, the presence of a limited efficiency also affects the reconstruction fidelity as it further smoothens the posterior reconstruction around $N_{true}$. The results in Figs 4a and 4b show systematic ordering with the fidelity reconstruction of 4-state system greater than that of 3-state systems and this one greater than that of 2-state systems, which confirms that increasing photon-number resolution improves the Bayesian reconstruction. Noticeably, the 4-state systems with $D=4$ overperform both the 3- and the 2-state devices with $D=8$, highlighting the advantage of a higher-state system over a certain increase of multiplexing.
In Section III we also anticipated that the prior distribution occurring in the Bayesian reconstruction Eq.16 may penalize the reconstruction fidelity when the prior distribution attributes low probability to certain true photon-numbers. The prior poissonian distributions used for Fig.4a and 4b attributes smaller probabilities to photon-numbers which are further away from $\mu$, which is another penalizing factor for the reconstruction fidelity of large true photon-numbers.

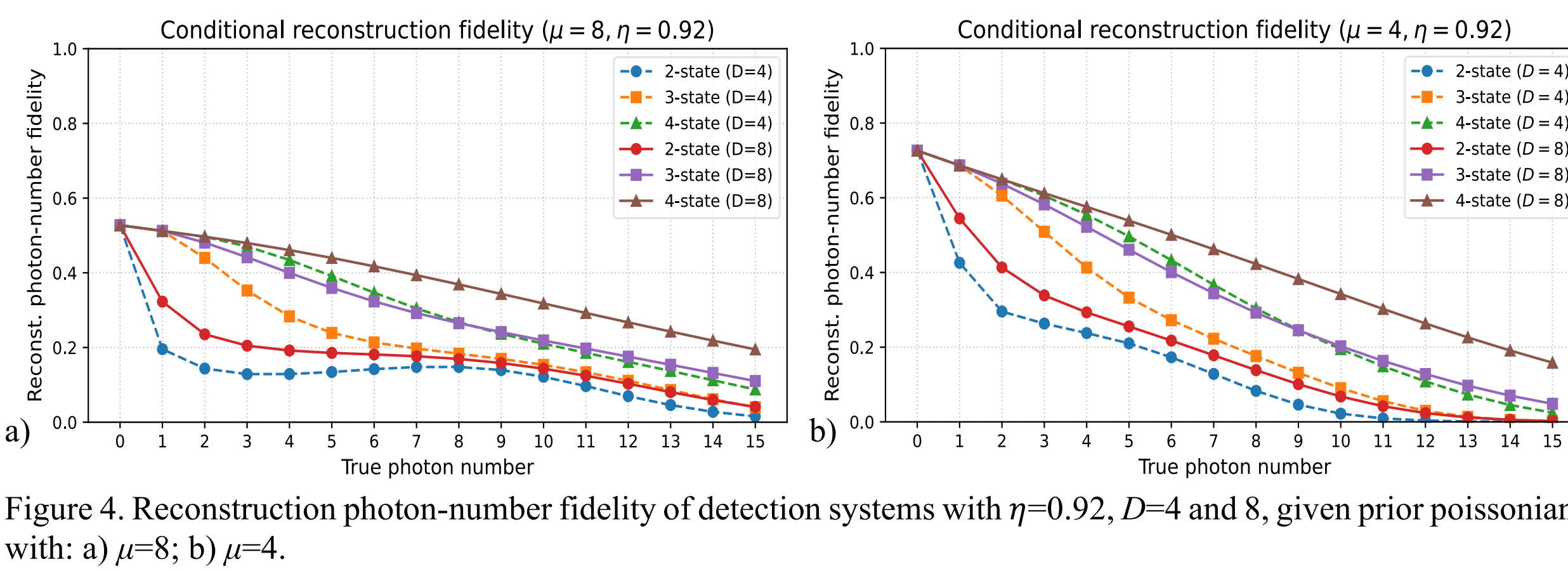

Figure 4. Reconstruction photon-number fidelity of detection systems with $\eta=0.92$, $D=4$ and 8, given prior poissonians with: a) $\mu=8$; b) $\mu=4$.

On the contrary, when the true photon-number is close to average number of photons of the light source, the reconstruction fidelity undergoes smaller degradation, and this effect is more evident for low state-systems in both Figs 4a and 4b, where the fidelity is lower.
Moreover, the reconstruction fidelities in Figure 4b, where $\mu$=4, are generally greater than the ones in Figure 4a, where $\mu$=8, because greater values of the posterior distribution are now found around lower values of true photon-number, where obviously the detection devices exhibit better photon-number resolving capabilities.

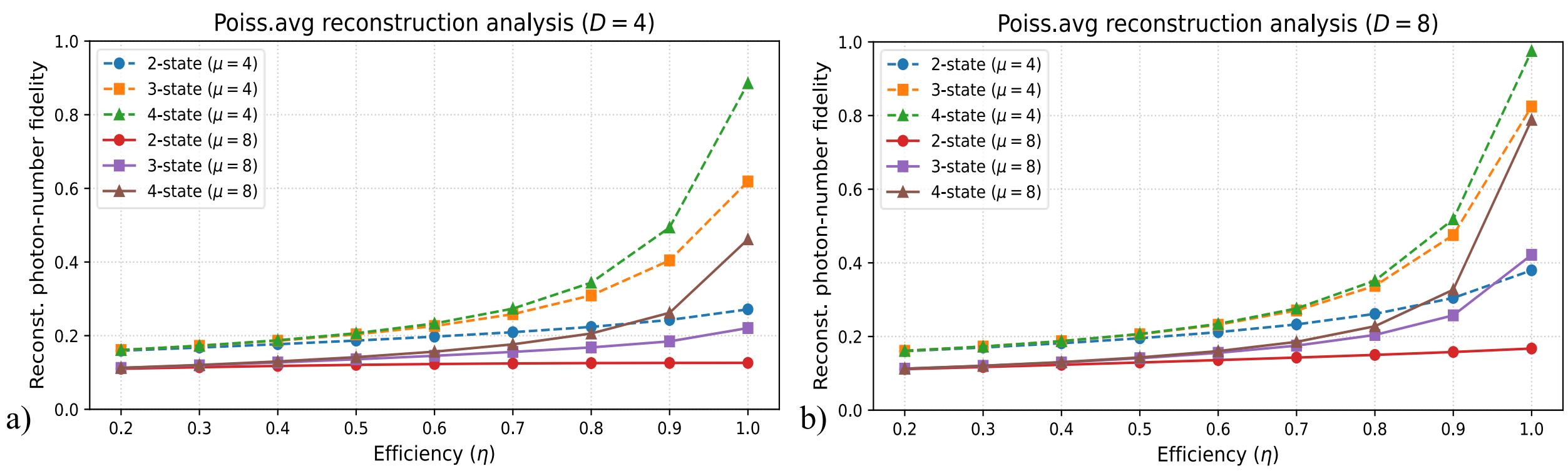

Figure 5. Poisson-averaged reconstruction analysis as a function of the detector efficiency for: a) systems with $D$=4; b) systems with $D$=8.

In Figure 5, the reconstruction photon-number fidelities at different true photon-numbers are averaged over all possible true photon-numbers that follow the same statistical distribution as the source, as per Eq.25. Evidently, now the result in terms of Poisson averaged reconstruction fidelity is a single value for the whole range of true photon-number. For this reason, the poisson-averaged reconstruction fidelities, expressed as functions of the detector efficiency, are smaller than 0.6 for $\eta$< 0.9. However, the fidelities largely increase for greater values of $\eta$, especially for cases with $D$=8 in Figure 5b. This is particularly advantageous in case of multiplexed, multi-state SNSPD devices, because the SNSPD can easily exceed $\eta = 0.9$. Also in this case, smaller $\mu$ of the prior provides greater efficiency for the same reasons seen above, but the results for 4-state detectors with μ=8 overperform the ones for 2-state with $\mu$=4.
A closing remark is worth noting on why both the $N$-photon detection fidelity and the reconstruction fidelity (in its global or prior-averaged form) matter, since much emphasis in characterizing the performance of photon-number resolving detectors is traditionally placed on the first one. $N$-photon detection fidelity expresses whether or not the detector output matches the true photon-number, which is undoubtedly useful information. However, the reconstruction fidelity answers another question, namely how much information about the true photon-number is gained from the detector output, and it is especially useful when the photon-number estimate is used to inform a further decision. In fact, as seen in this Section, a saturated outcome that matches the true photon-number does not degrade the detector fidelity, while the posterior it produces remains spread over several photon-numbers and the reconstruction fidelity is reduced. Generally, any protocol that acts on a photon-number threshold is only reliable when, given the observed outcome, the residual probability mass over that threshold is sufficiently small. The reconstruction fidelity quantifies exactly this probability. For instance, in experiments triggered by the detection of a single photon produced in a heralded single-photon source, the reconstruction fidelity can be used to assess if the underlying posterior assigns appreciable weight to two or more photons, therefore quantifying the risk of hidden multi-photon contamination.

## V. CONCLUSIONS

We have developed a combinatorial framework for the analysis of detection of multiplexed systems constituted by detectors provided with any number of states and achieved two main conceptual goals: we have generalized the formulas found in literature, [18], for multiplexed systems based only on two-state (on/off) detectors; we have shown that systems constituted by detectors with higher number of states, thus provided with intrinsic photon-number resolution, provide a quantifiable advantage over traditional solutions. This approach was recently stated as the next-stage development in improving photon-number resolution of detection systems [17,23].

The general formulation proposed consists of closed-form ideal and lossy photon detection probability tensors, obtained through combinatorics, for the multi-state multiplexed system with a given quantum efficiency. The formulation also includes the reconstruction of the photon-number distribution of the source from the detector output, obtained by means of Bayesian inference. To quantify the informative content of multi-state multiplexed systems we have derived formulas for the *N*-photon detection fidelity, by using photon detection probability tensors, and the reconstruction fidelity, based on the Bayesian posterior distribution. Accounting for dark counts noise only alters the expression of the photon detection probability tensor.
The informative advantage of multi-state multiplexed systems over the on/off systems is already evident with 3-state detectors, as they make it possible to relax the quadratic scaling law of traditional multiplexed systems, $D \sim N^2$ stated in [2]. The reconstruction fidelity confirms that this advantage persists, with a 4-state system with only *D*=4 elements consistently outperforming 2- and 3-state systems with twice as many detectors. Reconstruction fidelities remain moderate across all configurations considered, reflecting the intrinsic difficulty of unambiguously attributing true photon-numbers when saturation and losses are both present. Poisson-averaged results show that, using higher number-state arrays of detectors allows reaching high-fidelity photon-number resolution in a much more efficient way than scaling up a purely binary array.
Future studies should include experimental implementation of multi-state multiplexed systems, as single SNSPDs have already shown intrinsic photon-number resolution properties.

## ACKNOWLEDGEMENTS

This work was supported by the PN RIC 2021–2027 project 'Q-SUD - Polo meridionale di innovazione per l'informazione quantistica (identification code: QIIR112_00051), co-funded by the EU (ERDF) and the Italian Ministry (MUR), CUP: B49H26000300007.

## APPENDIX: Accounting for dark counts

In principle, the dark count noise rate of an SNSPD coincides with the signal rate of the device measured when no radiation reaches the detector. Practically, the measurement is simply done by switching off the light source and assuming that all counts derive from device noise. By suppose that the occurrence of dark counts follows a poissonian statistics, the probability of false positives is

$$\epsilon = 1 - e^{-R_{DC}\tau} \approx R_{DC}\tau \tag{A1}$$

Where $R_{DC}$ is the dark count rate measured, $\tau$ is the time window defined for a detection event. $\tau$ represents the time interval in which a click is attributed to a light impulse: it does not depend on the device, and it is not universally defined. For Hz- dark count rates and a wide range of $\tau$ (from picoseconds to microseconds), the linear approximation for $\epsilon$ provides negligible errors. Using $\tau$ as the coincidence window of ref [17] (~300 ps) may provide widely different results, in terms of $\epsilon$, from taking $\tau$ as the time period between two impulses of a laser source (e.g. ~100 µs for a pulse-picked laser impulse at 10 kHz frequency). Using $R_{DC} = 1\ Hz$ in both cases, and one can get $\epsilon$~3E-10 for the former, and $\epsilon$~1E-4 for the latter example.
In this paper we follow ref. [13] by assuming $\tau$ in the range of the resistive domain lifetime (50-800 ps), which directly assesses the possibility of a dark-count-generated hotspot to combine with a photon-generated-hotspot before this one collapses. In this case, the effect of dark counts on *N*-photon detector fidelity and reconstruction fidelity is indeed negligible.
In the main paper, we have proposed a mathematical formulation for the response tensor, Eq.12, where the dark counts are not accounted for, and we have shown numerical results that use this noiseless response tensor. However, to cover also situations where the number of photons is extremely low, for instance in some single photon sources and heralded experiments, and $\epsilon$ becomes significant due to longer $\tau$, we include in this appendix the formulation for the response tensor that includes the presence of dark counts. The final expression heavily complicates, but the formulas for the *N*-photon detection fidelity and the reconstruction fidelities proposed in the main paper remain unchanged.
To deal with the problem of dark counts, we make several assumptions (hypotheses) based on known facts about dark counts in SNSPDs:
1. Each detector can produce a dark count with the same probability $\varepsilon$ in the time window considered.

2. We assume that maximum one dark count can be generated on each detector.
3. Each dark count contributes to the counting process in the same way as a photon does, since it generates identical nucleation/growth hotspot dynamics, as suggested in ref [30].
Hp.1 is in line with the assumptions made in ref [18] for on/off multiplexed detectors.
Hp.2 is motivated by merely probabilistic reasons, as the probability of generating more than one dark count on a nanowire is extremely low. However, there seems no physical reason that would forbid the presence of two or more dark counts on a nanowire in a given time window.
Hp.3 is a consequence of well-known facts from the literature [30]: both the arrival of a photon on a SNSPD in its superconductive state and the spontaneous event generating a dark count trigger the same sequence of events: hotspot nucleation, hotspot expansion and diversion of superconducting current, resulting in a count. This means that a noise-generated and a photon-generated count should be indistinguishable, as they produce the same local suppression of the superconducting order parameter.
Hp.3 has strong implications in case of more-than-two-state detectors: the presence of a dark-count will be resolved in the same way as that of a photon, therefore, when a dark count appears in one detector belonging to a certain detection class $k_r$, the detector leaves this class and moves to the next upper one, $k_{r+1}$. However, if the dark count occurs in one of the detectors of the saturated class, $k_{s-1}$, no detection takes place, leaving the output vector unchanged. In general, if $\boldsymbol{d} = (d_0, d_1, \dots, d_{s-1})$ dark counts occur on detectors belonging to the "pre-dark count" output vector $\boldsymbol{K}^{(s)}$, this one modifies into:

$$\boldsymbol{K}^{(s)} \overset{\boldsymbol{d}}{\rightarrow} \boldsymbol{K}'^{(s)}\colon \Phi^{(s)}\left(\boldsymbol{K}^{(s)}, d\right) = \left(k_0'^{(s)}, k_1'^{(s)}, \dots, k_{s-1}'^{(s)}\right) = \left(k_0^{(s)} - d_0, k_1^{(s)} - d_1 + d_0, \dots, k_{s-2}^{(s)} - d_{s-2} + d_{s-3}, k_{s-1}^{(s)} + d_{s-2}\right) \tag{A2}$$

The constraint on $\boldsymbol{d}$ is that the components of the resulting $\boldsymbol{K}'^{(s)}$ must sum up to *D*. Notice that the dark counts $d_{s-1}$ occurring in the saturated class $k_{s-1}^{(s)}$ leave the class unchanged and produce no effects. Therefore, we now assume that the dark count vector is provided with only the first *s*-2 components, $\boldsymbol{d} = (d_0, d_1, \dots, d_{s-2})$. The equation remains unchanged.
Transforming $\boldsymbol{K}^{(s)}$, generated with Eq.12 thus accounting for losses, into $\boldsymbol{K}'^{(s)}$, that accounts for the presence of dark counts, occurs with a certain probability, that we name output vector transformation probability tensor, $T_\epsilon^{(s)}\left(\boldsymbol{K}'^{(s)} | \boldsymbol{K}^{(s)}\right)$. Deriving this tensor would directly yield the probability response tensor of the detector by composing these two stochastic processes without mixing them [26], as:

$$P\left(\boldsymbol{K}'^{(s)} | N, D, \eta, \epsilon\right) \sum_{\boldsymbol{K}} T_\epsilon^{(s)}\left(\boldsymbol{K}'^{(s)} | \boldsymbol{K}^{(s)}\right) P\left(\boldsymbol{K}^{(s)} | N, D, \eta\right) \tag{A3}$$

The summation constrained over $\boldsymbol{K}$ means that all the pre-dark counts output vector space must be explored, i.e. $k_0, k_1, \dots, k_{s-1} \geq 0$ and $k_0 + k_1 + \dots + k_{s-1} = D$. $T_\epsilon^{(s)}\left(\boldsymbol{K}'^{(s)} | \boldsymbol{K}^{(s)}\right)$ can be derived considering that, given $\varepsilon$, the probability that $d_i$ dark counts occur in $k_i$ detectors is given by the binomial convolution:

$$P_{dc}(d_i | k_i, \epsilon) = \binom{k_i}{d_i} \epsilon^{d_i} (1 - \epsilon)^{k_i - d_i} \tag{A4}$$

To account for the probability of having the whole dark count vector $\boldsymbol{d}$ occurring over all the classes of the detection system, we have to multiply each term obtained with *i*=0,1,…*s*-1:

$$P_{dc}\left(\boldsymbol{d} | \boldsymbol{K}^{(s)}, \epsilon\right) = \prod_{i=0}^{s-2} P_{dc}(d_i | k_i, \epsilon) \tag{A5}$$

From $P_{dc}\left(\boldsymbol{d} | \boldsymbol{K}^{(s)}, \epsilon\right)$, the next step to $T_\epsilon^{(s)}\left(\boldsymbol{K}'^{(s)} | \boldsymbol{K}^{(s)}\right)$ is linking $\boldsymbol{d}$ to $\boldsymbol{K}'^{(s)}$, which requires that the whole space of vector $\boldsymbol{d}$ is included and the $\boldsymbol{K}^{(s)} \overset{\boldsymbol{d}}{\rightarrow} \boldsymbol{K}'^{(s)}$ relations are respected:

$$T_\epsilon^{(s)}\left(\boldsymbol{K}'^{(s)}|\boldsymbol{K}^{(s)}\right) = \sum_{\boldsymbol{d}} P_{dc}\left(\boldsymbol{d}|\boldsymbol{K}^{(s)},\epsilon\right)\ \ with\ constraints\ from\ Eq.A2 \qquad \text{(A6)}$$

Summation over the vector space $\boldsymbol{d}$ means that all possible dark count vectors that fulfill Eq.A2 must be considered. Practically, considering Hp.3, this would imply $d_0, d_1, \dots, d_{s-2} \geq 0$ and $d_0 \leq k_0, d_1 \leq k_1, \dots, d_{s-2} \leq k_{s-2}$.
Using Eq.s A6 and A5, Eq.A3 becomes:

$$P\left(\boldsymbol{K}'^{(s)}|N,D,\eta,\epsilon\right) = \sum_{\boldsymbol{K}}\sum_{\boldsymbol{d}} P_{dc}\left(\boldsymbol{d}|\boldsymbol{K}^{(s)},\epsilon\right)\ P\left(\boldsymbol{K}^{(s)}|N,D,\eta\right)\ with\ constraints\ from\ Eq.A2 \qquad \text{(A7)}$$

Eq. A2, expressing $\boldsymbol{K}^{(s)} \xrightarrow{\boldsymbol{d}} \boldsymbol{K}'^{(s)}$, can be rewritten by taking $\boldsymbol{K}^{(s)}$ or $\boldsymbol{d}$ at the right-hand side and each vector can be unequivocally computed from the remaining two (three-way bijection). Therefore, exploring both vector spaces of $\boldsymbol{K}^{(s)}$ and $\boldsymbol{d}$, as indicated by the two summations in Eq. A7, leads to either negative components of any di component of the vector $\boldsymbol{d}$, or to promoting from a class $k_i^{(s)}$ to $k_{i+1}^{(s)}$ more detectors than the ones existing in the class $k_i^{(s)}$. In both cases, the corresponding probability is zero. Therefore, we can conclude that Eq. A7 can be written as:

$$P\left(\boldsymbol{K}'^{(s)}|N,D,\eta,\epsilon\right) = \sum_{\boldsymbol{d}} P_{dc}\left(\boldsymbol{d}|\boldsymbol{K}^{(s)},\epsilon\right)\ P\left(\boldsymbol{K}^{(s)}|N,D,\eta\right)\ with\ constraints\ from\ Eq.A2 \qquad (A8)$$

It is possible to verify that, for $\epsilon$=0, Eq. A8 reduces to Eq.12 in the main paper and, for $s$=2, to the theorem for noise detection of on/off multiplexed systems studied in ref [18].
Finally, the constraints from Eq. A2 can be explicitly expressed by using the Kronecker delta and the equation becomes:

$$P\left(\boldsymbol{K}'^{(s)}|N,D,\eta,\epsilon\right) =$$
$$= \sum_{\boldsymbol{d}}\left[\prod_{i=0}^{s-2}\binom{k_i}{d_i}\epsilon^{d_i}(1-\epsilon)^{k_i-d_i}\right]P\left(\boldsymbol{K}^{(s)}|N,D,\eta\right)\delta_{k'_0,k_0-d_0}\delta_{k'_1,k_1-d_1+d_0}\dots\delta_{k'_{s-2},k_{s-2}-d_{s-2}+d_{s-3}}\delta_{k'_{s-1},k_{s-1}+d_{s-2}} \quad (A9)$$

The $P\left(\boldsymbol{K}^{(s)}|N,D,\eta\right)$ derives from Eq. 12 in the main paper.